\documentclass[prd,aps,a4paper,preprint,preprintnumbers,nofootinbib]{revtex4-1}
\usepackage[a4paper, hdivide={1.91cm,,1.165cm}, vdivide={1.83cm,,3.3cm}]{geometry}

\usepackage{amsmath,amssymb}
\usepackage{graphicx,multirow}
\usepackage{tabularx}
\usepackage[export]{adjustbox}
\usepackage[hyperfootnotes=false]{hyperref}
\usepackage[usenames, dvipsnames]{color}
\usepackage[normalem]{ulem}
\usepackage{float}

\newcommand{\ii}{\ensuremath{\text{i}}}
\newcommand{\dd}{\ensuremath{\text{d}}}

\newcommand{\beq}{\begin{eqnarray}}% can be used as {equation} or  {eqnarray}
\newcommand{\eeq}{\end{eqnarray}}

\begin{document}
%%%%%%%%%%%%%%%%%%%%%%%%%%%%%%%%%

\preprint{UCI-TR-2021-10}

\title{Mapping Gauged Q-Balls}

\author{Julian~Heeck}
\email{heeck@virginia.edu}
\affiliation{Department of Physics, University of Virginia,
Charlottesville, Virginia 22904-4714, USA}

\author{Arvind~Rajaraman}
\email{arajaram@uci.edu}
\affiliation{Department of Physics and Astronomy, 
University of California, Irvine, CA 92697-4575, USA
}

\author{Rebecca~Riley}
\email{rebecca.riley@uci.edu}
\affiliation{Department of Physics and Astronomy, 
University of California, Irvine, CA 92697-4575, USA
}

\author{Christopher~B.~Verhaaren}
\email{cverhaar@uci.edu}
\affiliation{Department of Physics and Astronomy, 
University of California, Irvine, CA 92697-4575, USA
}

%\date{\today}

\begin{abstract}
Scalar field theories with particular $U(1)$-symmetric potentials contain non-topological soliton solutions called Q-balls. Promoting the $U(1)$ to a gauge symmetry leads to the more complicated situation of gauged Q-balls. 
The soliton solutions to the resulting set of nonlinear differential equations have markedly different properties, such as a maximal possible size and charge. Despite these 
differences, we discover a relation that allows one to extract the properties of gauged Q-balls (such as the radius, charge, and energy) from the more easily obtained properties of global Q-balls. These results provide a new guide to understanding gauged Q-balls as well as providing simple and accurate analytical characterization of the Q-ball properties.
\end{abstract}
%%%%%%%%%%%%%%%%%%%%%%%%%%%%%%%%%
%%%%%%%%%%%%%%%%%%%%%%%%%%%%%%%%%
\maketitle

\tableofcontents
%%%%%%%%%%%

\section{Introduction}

Q-balls are stable nontopological solitons that can arise in theories 
involving complex scalars $\phi$~\cite{Coleman:1985ki} (for a review see Ref.~\cite{Nugaev:2019vru}). In 
the case of global 
Q-balls, $\phi$ carries a conserved global charge and the solitons are stabilized by a scalar potential 
that provides an attractive force~\cite{Lee:1991ax}. 
Global Q-balls have been proposed as dark matter~\cite{Kusenko:1997si,Kusenko:2001vu} due to 
their potential occurrence in supersymmetric models and provide in particular a simple realization of macroscopic 
dark matter~\cite{Ponton:2019hux,Bai:2020jfm}. 

The analytic construction of Q-balls requires solving a nonlinear differential equation.
In certain potentials, the equation can be solved exactly~\cite{Rosen:1969ay,Theodorakis:2000bz,MacKenzie:2001av,Gulamov:2013ema}.
For many other cases numerical solutions can be efficiently
obtained via computer programs such as \texttt{AnyBubble}~\cite{Masoumi:2016wot}.
Recently, it was shown that almost all aspects of global Q-balls can be 
understood essentially
 analytically, even for potentials which are not exactly solvable~\cite{Heeck:2020bau}. Extremely accurate 
analytical expressions
were obtained for global Q-ball properties such as radius, charge, and energy in some non-solvable scenarios which 
essentially obviate the need for numerical studies~\cite{Heeck:2020bau}. 
It appears that for all intents and purposes single-field global Q-balls are a solved problem.

The system's complexity increases if $\phi$ is charged under a local symmetry, which 
 leads to \emph{gauged} Q-balls~\cite{Lee:1988ag,Gulamov:2015fya,Brihaye:2015veu}. 
Given the prevalence of gauge bosons in the Standard Model and its extensions, understanding gauged 
Q-balls is important phenomenologically. However, they are considerably more difficult to 
describe, both analytically and numerically. On the analytic side, no exactly solvable 
examples are known to us. Numerical studies are made difficult by the gauge field, which 
appears in the scalar potential as a field whose kinetic term has the opposite sign. This makes numerical 
studies (using, for example, the ever-popular shooting method) far more tedious to implement. 

In this article we extend the methods of Ref.~\cite{Heeck:2020bau} to gauged Q-balls. In so doing we reveal a close connection between global Q-balls and gauged Q-balls. This enables 
us to use our understanding of global Q-balls to analytically calculate
 the  properties of these gauged Q-balls\textemdash such as radius, charge, and energy. 
Furthermore, we find simple expressions for the scalar and gauge-field profiles that can be 
used to solve the differential equations efficiently using finite-element methods. This work paves 
the way for detailed phenomenological studies of these objects.

In the next section, 
we review global Q-balls and establish our notation.
Section~\ref{sec:gauged} introduces gauged Q-balls and analytical approximations for the scalar and 
gauge field profiles. 
In Sec.~\ref{sec:numerics} we present a method for solving the Q-ball differential equations 
using finite-element methods rather than the shooting method.
The novel mapping between global and gauged solutions is given in Sec.~\ref{sec:Mapping}.
The accuracy of our analytical predictions for the Q-ball profiles and 
observables such as energy, mass, and radius, are established in Sec.~\ref{sec:results}. We also derive quantities of interest such as the parametric regions of Q-ball 
stability before concluding in Sec.~\ref{sec:conclusion}.
A derivation of the Q-ball energy and alternative derivation of the mapping formula is given in Appendices~\ref{a.Energy} and~\ref{sec:alternative_derivation}, respectively.

%%%%%%%%%%%%%%%%%%%%
%%%%%%%%%%%%%%%%%%%%
\section{Review of Global Q-Balls}
\label{sec:global}

The Lagrangian density   for a complex scalar $\phi$ 
\begin{equation}
\mathcal{L}=\left|\partial_\mu\phi \right|^2-U(|\phi|),
\end{equation}
 enjoys an explicit global $U(1)$ symmetry $\phi\to e^{\ii \alpha} \phi$. The conserved charge 
$Q$ under this symmetry is $\phi$ number, normalized so that $Q(\phi)=1$.
To preserve the $U(1)$ symmetry, we require $\langle\phi\rangle=0$ in the vacuum. We choose the potential 
energy to be zero in the vacuum 
by setting $U(0)=0$ and enforce that the vacuum is a stable minimum of the potential by 
\begin{equation}
\left.\frac{\dd U}{\dd |\phi|}\right|_{\phi=0}=0~, \ \ \ \ \left.\frac{\dd^2 U}
{\dd\phi\, \dd\phi^\ast}\right|_{\phi=0}=m_\phi^2\,,
\end{equation}
where $m_\phi$ is the mass of the complex scalar. 
In this scenario, Coleman~\cite{Coleman:1985ki} showed that nontopological solitons, Q-balls, exist when the 
function $U(|\phi|)/|\phi|^2$ has a minimum at $|\phi|=\phi_0/\sqrt{2}>0$ such that 
\begin{equation}
0\leq \frac{\sqrt{2U(\phi_0/\sqrt{2})}}{\phi_0}\equiv\omega_0<m_\phi\,.\label{e.Omega0}
\end{equation}
Spherical Q-ball solutions have the form
\begin{equation}
\phi (t,\vec{x})=\frac{\phi_0}{\sqrt{2}} f(r) e^{\ii\,\omega_G t}\,,\label{e.fdef}
\end{equation}
for a constant $\omega_0<\omega_G < m_\phi$. We choose $\omega_G$ to be positive, which results in a positive 
charge $Q$ of the Q-ball.
It is convenient to define the dimensionless quantities 
\begin{align}
\rho\equiv r\sqrt{m_\phi^2-\omega_0^2}\,, && 
\Omega_G\equiv\frac{\omega_G}{\sqrt{m_\phi^2-\omega_0^2}}\,, &&
\Omega_0\equiv\frac{\omega_0}{\sqrt{m_\phi^2-\omega_0^2}}\,, &&
\Phi_0\equiv\frac{\phi_0}{\sqrt{m_\phi^2-\omega_0^2}}\,.
\label{eq:rescaling}
\end{align}
We can then write the Lagrangian as
\beq
L=4\pi\Phi_0^2\sqrt{m_\phi^2-\omega_0^2}\int \dd\rho\; \rho^2\left[-\frac12f^{\prime2}+\frac12f^2\Omega_G^2-
\frac{U(f)}{\Phi_0^2(m_\phi^2-\omega_0^2)^2} \right] ,
\eeq
where a prime denotes a derivative with respect to $\rho$. The equation of motion for $f$ is 
\beq
f'' + \frac{2}{\rho} f'=\frac{1}{\Phi_0^2(m_\phi^2-\omega_0^2)^2}\frac{\dd U}{\dd f}-\Omega_G^2f\,.\label{e.fGlobal}
\eeq
Q-ball solutions for $f$ satisfy this nonlinear differential equation along with the boundary conditions 
$f'(\rho\to 0)=0 = f(\rho\to\infty)$.

As an explicit example, we consider the most generic $U(1)$-symmetric sextic potential
studied in Ref.~\cite{Heeck:2020bau}. This can be parametrized as
\begin{align}
U(f) = \phi_0^2 \left( \frac{m_\phi^2 -\omega_0^2}{2}\, f^2 (1-f^2)^2 +\frac{\omega_0^2}{2}\, f^2 \right) .
\label{eq:sextic_potential}
\end{align}
The differential equation of Eq.~\eqref{e.fGlobal} then takes the form
\begin{align}
f'' + \frac{2}{\rho} f'= f \left( 1-\kappa_G^2 -4 f^2  +3 f^4 \right) ,
\end{align}
where  $\kappa_G^2 \equiv \Omega_G^2 - \Omega_0^2$. The solutions depend on the single parameter 
$\kappa_G\in (0,1)$, which also determines
 the (dimensionless) Q-ball radius $R^\ast$.\footnote{The definition of $R^\ast$ is 
somewhat ambiguous as $f$ transitions smoothly from its value at the center of the 
Q-ball to its value outside, but a useful definition is  $f''(\rho=R^\ast)=0$. 
}

For small $\kappa_G$, the Q-balls are large and the relation $R^\ast(\kappa_G)$ can
be calculated analytically at leading order to be
$R^\ast(\kappa_G)=1/\kappa_G^2$~\cite{Heeck:2020bau}.
For these large Q-balls,
the exact Q-ball profile is close to a step function 
$f(\rho) \simeq 1 - \Theta(\rho-R^\ast)$, this is the so-called \emph{thin wall} 
limit~\cite{Coleman:1985ki}.
As shown in Ref.~\cite{Heeck:2020bau}, an even better profile for these {thin-wall} 
Q-balls around $\rho\sim R^\ast \gg 1$ is
\begin{align}
f_T(\rho)=\frac{1}{\sqrt{1+2e^{2(\rho-R^\ast)}}} \,.
\label{eq:transition_profile}
\end{align}
This is called the transition profile, since it describes the rapid transition from the nearly 
constant $f\simeq 1$ inside the Q-ball to $f\simeq 0$ outside the Q-ball.
The transition profile is actually a very good approximation to the full profile
 for all $\rho$ and even works reasonably well for smaller Q-balls~\cite{Heeck:2020bau}.
 
We also present  here a new relation for $R^\ast(\kappa_G)$ 
\begin{align}
R^\ast(\kappa_G)=\frac{1}{\kappa_G^2}-\frac{1}{4\kappa_G}+\frac32-2\kappa_G+\frac{1}{3\sqrt{1-\kappa_G^2}}~,
\label{e.globalRapproximation}
\end{align}
which provides an approximation to the numerical result that is accurate to better than 
2\% in the region $\kappa_G < 0.84$ (or $R^\ast\gtrsim 1.5$) that leads to 
 stable Q-balls (i.e.  Q-balls with $E < m_\phi Q$).
This relation can be used
to produce extremely accurate expressions of the global Q-ball's energy and charge as a function of radius 
using the expressions in Ref.~\cite{Heeck:2020bau}.

\begin{figure}[t]
\includegraphics[width=0.6\textwidth]{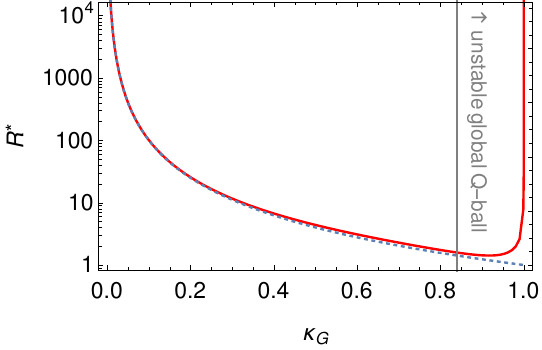}
\caption{
The global Q-ball radius $R^\ast$ vs $\kappa_G = \sqrt{\Omega_G^2 - \Omega_0^2}$ for the sextic 
potential~\cite{Heeck:2020bau} in red. The blue dotted line shows the  approximation $\kappa_G^2 = 1/R^\ast$. The region 
$\kappa_G \gtrsim 0.84$ leads to unstable global Q-balls due to $E> m_\phi Q$~\cite{Heeck:2020bau}.
}
\label{fig:kappaRglobal}
\end{figure}

%%%%%%%%%%%%%%%%%%%%%%%%%
%%%%%%%%%%%%%%%%%%%%%%%%%
\section{Gauged Q-Balls}
\label{sec:gauged}
 
Gauged Q-balls result from promoting the global $U(1)$ symmetry to a local symmetry.
The Lagrangian density is
\begin{equation}
\mathcal{L}=\left|D_\mu\phi \right|^2-U(|\phi|)-\frac14 F_{\mu\nu}F^{\mu\nu},
\end{equation}
where $D_\mu=\partial_\mu-\ii e A_\mu$ is the gauge covariant derivative and $F_{\mu\nu}=\partial_\mu A_\nu
-\partial_\nu A_\mu$ is the field-strength tensor. The parameter $e$ is the gauge coupling normalized so that 
$\phi$ has charge one. 
After making the static charge ansatz~\cite{Lee:1988ag} 
\begin{equation}
\phi (t,\vec{x})=\frac{\phi_0}{\sqrt{2}} f(r) e^{\ii\,\omega t}\,,\qquad A_0(t,\vec{x})=A_0(r)\,,\qquad A_i(t,\vec{x})=0\,,\label{e.fdef2}
\end{equation}
and defining dimensionless quantities 
\begin{align}
\Omega\equiv\frac{\omega}{\sqrt{m_\phi^2-\omega_0^2}}\,, \qquad
A(\rho)\equiv \frac{A_0 (\rho)}{\phi_0}\,,\qquad
 \alpha\equiv e\Phi_0\,,\qquad
\kappa^2 \equiv \Omega^2 - \Omega_0^2\,,
\label{eq:rescaling2}
\end{align}
we rewrite the Lagrangian as
\begin{equation}
L=4\pi\Phi_0^2\sqrt{m^2_\phi-\omega_0^2}\int \dd\rho\,\rho^2\left\{ -\frac12f^{\prime 2}+\frac{1}{2}A^{\prime2}
+\frac{1}{2}f^2\left(\Omega-\alpha A \right)^2-\frac{U(f)}{\Phi_0^2(m_\phi^2-\omega_0^2)^2}\right\}.\label{e.massesLag}
\end{equation}
This has the form of two scalar fields under the influence of the potential
 \beq
 V(f,A)=\frac{1}{2}f^2\left(\Omega-\alpha A \right)^2-\frac{U(f)}{\Phi_0^2(m_\phi^2-\omega_0^2)^2}\,.
 \label{eq:potentialV}
 \eeq
 However, it is important to notice that in this analogy the $A$ field's kinetic term has 
the wrong sign. 
The two equations of motion 
\begin{align}
f'' + \frac{2}{\rho} f'&= -\frac{\partial V}{\partial f} = \frac{1}{\Phi_0^2(m_\phi^2-\omega_0^2)^2}\frac{\dd U}
{\dd f}-\left(\Omega-\alpha A\right)^2f\,,\label{e.feq}\\
A'' + \frac{2}{\rho} A' &= +\frac{\partial V}{\partial A} = \alpha f^2(A\alpha-\Omega)\,,\label{e.Aeq}
\end{align}
are to be solved subject to the boundary conditions
\begin{equation}
\lim_{\rho\to0}f'= \lim_{\rho\to\infty}f =  \lim_{\rho\to0}A'= \lim_{\rho\to\infty}A =0 \,.
\end{equation}
In the analogy of two fields moving in the potential $V$, $\rho$ becomes a time coordinate and the terms with an explicit  $1/\rho$ 
can be interpreted as time-dependent friction terms. As shown below, this analogy greatly aids our understanding of the Q-ball solutions.

The scalar frequency $\omega$ is restricted to the region $\omega_0<\omega \leq m_\phi$; this 
is similar to the global Q-ball case, except that it is possible to have gauged Q-balls 
with $\omega = m_\phi$ (or $\kappa=1$)~\cite{Gulamov:2015fya}, where no global Q-balls exist. 
In section \ref{sec:results} we show that a stronger lower bound on $\omega$ exists.

The conserved charge $Q$ is defined in the usual way as the integral over the time component of the scalar 
current~\cite{Lee:1988ag}
\begin{align}
Q&=4\pi \Phi_0^2\int \dd\rho\,\rho^2f^2\left(\Omega-\alpha A \right) \label{e.charge}\\
&=-\frac{4\pi\Phi^2_0}{\alpha}\lim_{\rho\to\infty}\rho^2A' \,,
\end{align}
where the second line uses Eq.~\eqref{e.Aeq} and integration by parts.
This implies that for large $\rho$,
\begin{equation}
A=\frac{\alpha\, Q}{4\pi\,\Phi^2_0\,\rho},\label{e.asymG}
\end{equation}
up to corrections that fall off faster than $1/\rho$~\cite{Lee:1988ag}. 
The gauged Q-ball energy $E$ is  obtained from the Hamiltonian
\begin{align}
E/\sqrt{m_\phi^2-\omega_0^2} & =4\pi\Phi_0^2\int \dd\rho\,\rho^2\left\{ \frac12f^{\prime 2}+\frac{1}{2}A^{\prime2}
+\frac{1}{2}f^2\left(\Omega-\alpha A \right)^2+\frac{U(f)}{\Phi_0^2(m_\phi^2-\omega_0^2)^2}\right\}\label{e.Eint}\\
&= \Omega Q+\frac{4\pi\Phi_0^2}{3}\int \dd\rho\,\rho^2\left(f^{\prime 2}-A^{\prime2} \right) \label{e.energy}.
\end{align}
The second expression corrects a typo in Ref.~\cite{Lee:1988ag} and is derived in App.~\ref{a.Energy}.  
The energy and charge also satisfy the non-trivial differential equation~\cite{Gulamov:2013cra}
\beq
\frac{\dd E}{\dd\omega}=\omega\frac{\dd Q}{\dd\omega} \,.\label{e.dedw}
\eeq
This is a powerful relation among the Q-ball observables and, in particular, allows $\omega$ to be interpreted 
as the chemical potential.

For concreteness we restrict most of our discussion to the sextic scalar potential of 
Eq.~\eqref{eq:sextic_potential}, although we expect our results to be qualitatively 
applicable to a far larger class of potentials.
Just like in the global case we only study ground-state Q-balls, which have no nodes; excited gauged Q-balls 
in the same potential have been discussed in Ref.~\cite{Loginov:2020xoj}.

%%%%%%%%%%%%%%%%%%%%%
%%%%%%%%%%%%%%%%%%%%%%%
\section{Numerical Methods}
\label{sec:numerics}

While the shooting method is quite successful for global Q-balls~\cite{Coleman:1985ki}, the 
addition of the  gauge field makes finding a solution using this method tedious, especially for large 
Q-balls. We avoid this by changing coordinates and solving the boundary value problem 
directly. A similar approach was employed in Ref.~\cite{Panin:2016ooo}.

In order to enforce the boundary conditions at $\rho=\infty$ we switch to a 
compactified coordinate $y$,
\beq
y=\frac{\rho}{1+\rho/a}~,
\eeq
where $a$ is a positive constant. The value of $a$ makes no real difference in obtaining 
numerical solutions. However, choosing $a$ much larger than the Q-ball radius ensures that
the most drastic compactification effects occur outside the Q-ball. Clearly, $y$ takes values $y\in[0,a]$ and so we can  require the 
conditions $f(a)=0$ and $A(a)=0$. The derivatives become
\beq
\frac{\dd\,}{\dd\rho}=\frac{\dd y}{\dd\rho}\frac{\dd\,}{\dd y}=
\left(1-\frac{y}{a} \right)^2\frac{\dd\,}{\dd y}\,,
\eeq
so the boundary conditions at $y=0$ are  $f'(0)=0$ and $A'(0)=0$ where primes 
denote a derivative with respect to $y$. 
The set of equations
\begin{align}
&\left(1-\frac{y}{a} \right)^4\left(f''+\frac{2}{y}f' \right)+f\left(\kappa^2+\alpha A(\alpha A-2\Omega)
-1+4f^2-3f^4 \right)=0 \,,\\
&\left(1-\frac{y}{a} \right)^4\left(A''+\frac{2}{y}A' \right)-\alpha f^2\left(\alpha A-\Omega \right)=0\,,
\end{align}
can then be solved by finite element methods, using \texttt{Mathematica}'s~\cite{Mathematica} 
 routines for instance, and quickly converges to the exact solution
if the initial guess is reasonably accurate. 
In the next section we present a method for finding analytical test functions 
for $f$ and $A$ that 
are  close to the exact solutions. These can be successfully used as
initial seed functions for this method.

%%%%%%%%%%%%%%%%%%%%%
\section{Mapping Global Q-Balls to Gauged Q-Balls\label{sec:Mapping}}

Much of the Q-ball profile can be understood by comparing it to the motion of 
a particle moving in the potential of Eq.~\eqref{eq:potentialV}
\beq
V(f,A)=\frac{1}{2}f^2\left[\kappa^2+\alpha A(\alpha A-2\Omega)-\left(1-f^2\right)^2 \right] .
\eeq
 For constant $A$ the potential in $f\geq0$ has three extrema, one at $f=0$ and the other two at
\beq
f^2_\pm=\frac13\left(2\pm\sqrt{1+3\kappa^2-3\alpha A(2\Omega-\alpha A)} \right) ,\label{e.fpm}
\eeq
$f_+$ being a maximum and $f_-$ a minimum.

\begin{figure}[t]
\includegraphics[width=0.49\textwidth]{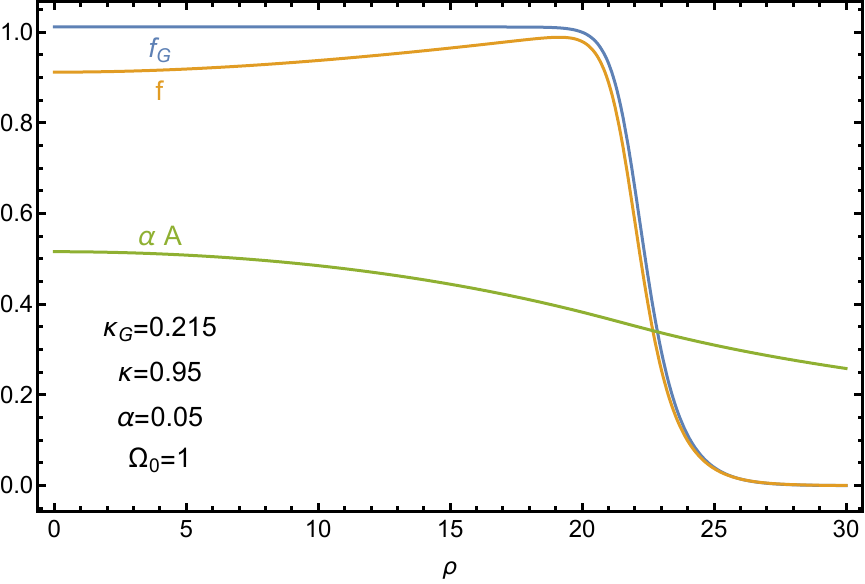}
\includegraphics[width=0.49\textwidth]{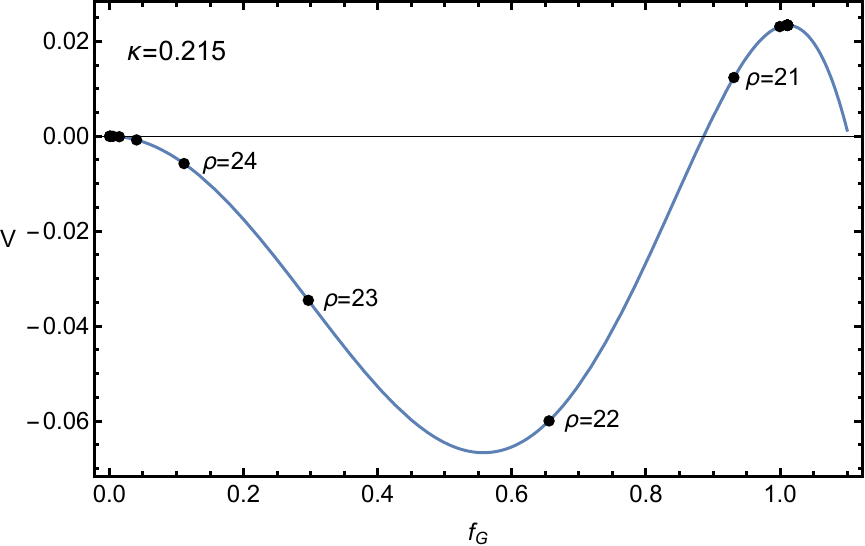}
\caption{
\emph{Left:} Profiles for global $f_G$ and gauged $f,\,\alpha A$ Q-balls corresponding to 
$R^\ast\approx 22$. \emph{Right:} 
Effective potential for the global Q-ball. Black points indicate the value of 
$f_G$ for integer values of $\rho\in[0,30]$. }
\label{fig:potentialPlot}
\end{figure}

For \emph{global} Q-balls, the second term in $V(f,0)$ vanishes; 
the scalar field starts close to the top of the
potential at $f\approx f_+(A=0)$. Eventually, the scalar rolls off and transitions
to the second maximum at $f=0$. Figure~\ref{fig:potentialPlot} gives an example global profile 
(blue curve of the left panel) along with the potential that determines its dynamics (right 
panel). Black points on the potential mark values of integer $\rho$, and illustrate that the 
field is nearly constant until $\rho\approx20$, after which the field rolls quickly.  The 
initial location of the field profile on the potential was found in Ref.~\cite{Heeck:2020bau} 
by 
matching the energy gap between the initial and final maxima to the loss of energy due 
to the friction-like term in the 
equation of motion.

\begin{figure}[t]
\includegraphics[width=0.49\textwidth]{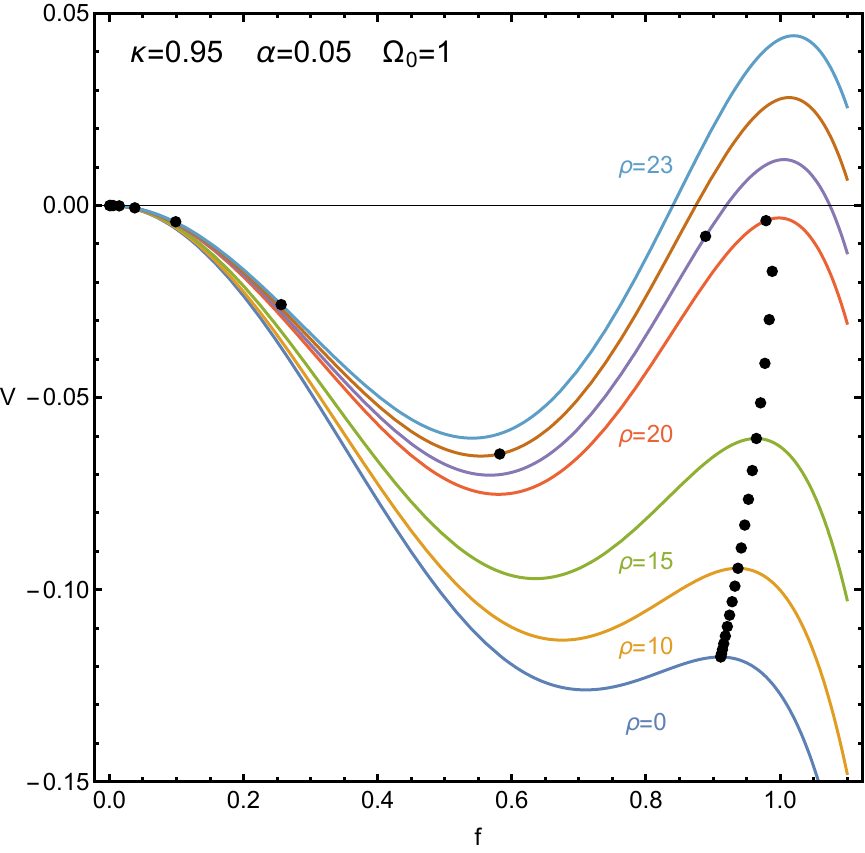}
\includegraphics[width=0.49\textwidth]{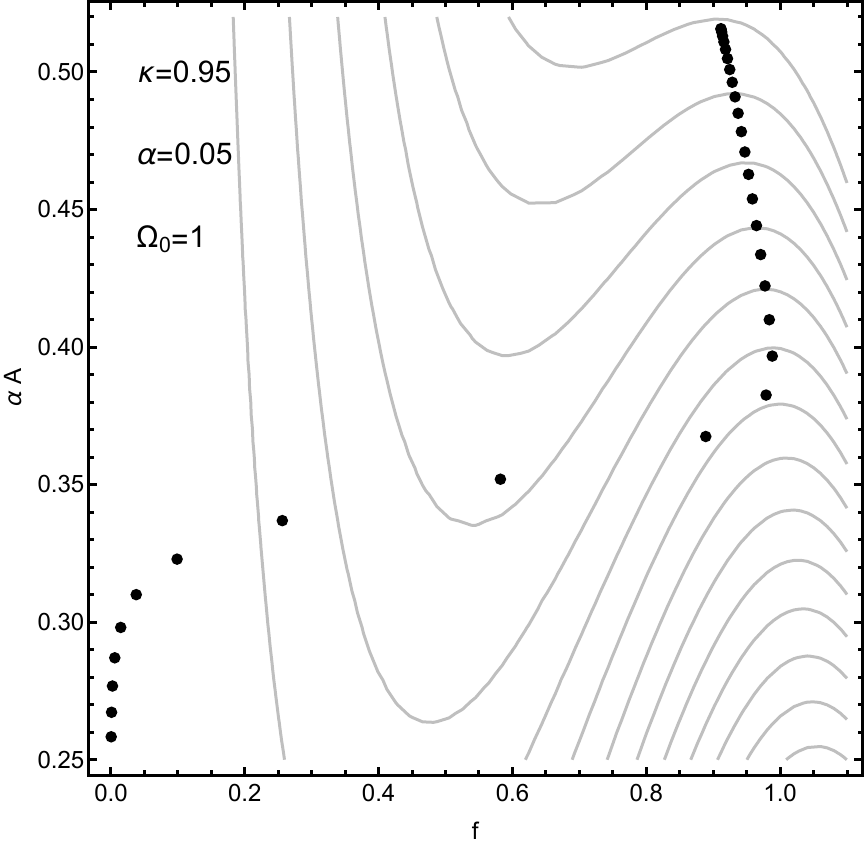}
\caption{Black points indicate the values of the gauged Q-ball profiles for integer values of $\rho\in[0,30]$. 
\emph{Left:} Effective potentials for $f$ given for specific values of $A(\rho)$. \emph{Right:} 
Contour plot of the potential $V$ as a function of $f$ and $\alpha A$. }
\label{fig:gaugepotentialPlot}
\end{figure}

Similar arguments apply to \emph{gauged} Q-balls.
 The primary difference between the
global and gauged cases is that the evolving gauge field $A$ causes the effective potential 
for the scalar to change with $\rho$, see the left panel of Fig.~\ref{fig:gaugepotentialPlot}.
The gauge field evolution changes the location and height of the second maximum at $f_+$,
 and the scalar continues to follow this maximum until a certain point when it
transitions quickly to the other maximum at $f=0$. Of course, this can only occur 
when $f_+$ exists, so the requirement that Eq.~\eqref{e.fpm} is real implies
\beq
\frac13+\kappa^2\geq\alpha A(2\Omega-\alpha A) \,.
\eeq
Notice that this condition is trivially satisfied in the global case, i.e.~for $\alpha\to 0$, but in the gauged 
case restricts $\alpha A$ to two possible regions:
\begin{align}
\alpha A\leq\Omega-\sqrt{\Omega_0^2-\frac13} \quad \text{ or } \quad \alpha A\geq\Omega+\sqrt{\Omega_0^2-\frac13} \,. 
\label{e.ArvindsInequality}
\end{align}
As shown below, the second inequality in Eq.~\eqref{e.ArvindsInequality} is not 
compatible with Q-ball solutions, leaving us with an \emph{upper} bound on $\alpha A$ when $\Omega_0\geq1/\sqrt{3}$.

As with the global case, we can determine the initial values of the fields by energy considerations.
Neglecting the  friction terms, we can write the equations of motion as 
\beq
f''+\frac{\partial V}{\partial f}=0, \ \ \ \ A''-\frac{\partial V}{\partial A}=0\,.
\eeq
This means that the quantity
\beq
\mathcal{E}=\frac12f^{\prime2}-\frac12A^{\prime2}+V(f,A)\,,
\eeq
is conserved as a function of $\rho$:
\beq
\frac{\dd\mathcal{E}}{\dd\rho}=f'\left(f''+\frac{\partial V}{\partial f} \right)-A'\left(A''-\frac{\partial V}
{\partial A} \right)=0\,.
\eeq
Of course, when the friction is included this quantity is not conserved and we immediately find that
\beq
\frac{\dd\mathcal{E}}{\dd\rho}=-\frac{2}{\rho}\left( f^{\prime2}-A^{\prime2}\right) .\label{e.WorkFric1}
\eeq
{ This justifies our interpretation of the}
 term on the right hand side of the equation as a friction. 

For constant $f$, the potential for $A$ has one extremum at
\beq
A_\text{max}=\frac{\Omega }{\alpha}  \,.\label{e.Amax}
\eeq
Again, Eqs.~\eqref{e.energy} and~\eqref{e.WorkFric1} indicate that $f'$ and $A'$ affect the 
energy differently. The $f$ profile behaves according to our usual intuition, but the $A$ 
kinetic term has the opposite sign. Consequently, as Eq.~\eqref{e.Amax} is a minimum in $V$, 
the dynamics of the system drive $A$ uphill either toward $A=0$ or $A\to\infty$. If $A$ is 
larger than $A_\text{max}$ it diverges as $\rho\to\infty$, which clearly does not satisfy the 
Q-ball boundary conditions. This implies that for Q-ball solutions $\Omega-\alpha A>0$, which 
has two consequences: First, because the right-hand side of the $A$ equation of motion
\beq
A''+\frac{2}{\rho}A'=- \alpha f^2(\Omega-A\alpha) \,,
\eeq
is always negative, $A$ is monotonically decreasing for Q-ball solutions~\cite{Lee:1988ag}. 
Second, as the system evolves the negative term under the square-root in Eq.~\eqref{e.fpm} 
becomes smaller so the value of $f_+$ grows. %, see Fig.~\ref{fig:gaugepotentialPlot}. 
For 
some solutions, such as the one shown in Fig.~\ref{fig:gaugepotentialPlot},  the ``force" from the $A$ 
gradient pushes $f$ uphill toward this growing $f_+$.

While the gauge field does affect the total Q-ball dynamics, it seems to play a relatively 
minor role when $f$ transitions from near one to near zero. This observation suggests a 
relationship between the global Q-ball solutions and gauged Q-ball solutions. To 
explore this, we need analytic expressions for $A$ and $f$. Beginning at the thin-wall 
limit, we approximate $f$ by a step function, $f(\rho) = 1 - \Theta(\rho-R^\ast)$ and then 
solve the equation of motion Eq.~\eqref{e.Aeq} for $A$. By demanding that $A(\rho)$ and 
its derivative be continuous at $\rho=R^*$ one finds~\cite{Lee:1988ag}
\beq
A(\rho)=\frac{\Omega }{\alpha}\left\{\begin{array}{cc}
\displaystyle 1-\frac{\sinh\left(\alpha\rho\right)}{\cosh\left(\alpha R^\ast \right)\alpha\rho}\,, 
& \rho<R^\ast\,,\\[0.4cm]
\displaystyle\frac{\alpha R^\ast-\tanh\left(\alpha R^\ast \right)}{\alpha\rho}\,, & \rho\geq R^\ast\,.\\
\end{array}\right. \label{e.thinA}
\eeq
Remarkably, this result is a good approximation to the exact gauge field solution even beyond the 
thin-wall regime.

This result indicates that the derivative of $\alpha A$ is small if the Q-ball radius $R^\ast$ is large:
\beq
|\alpha A'(R^\ast)|=\frac{\Omega}{R^{\ast}} \left|\frac{\tanh(\alpha R^\ast)-\alpha R^\ast}{\alpha R^\ast}\right| 
< \frac{\Omega}{R^{\ast}} \,,
\eeq
which implies that $\alpha A$ is essentially constant over the transition.
We can then refine our analysis of the scalar profile by solving the $f$ equation of motion around $\rho\sim R^*$ 
with a constant $A$:
\beq
f'' + \frac{2}{\rho} f'=\frac{1}{\Phi_0^2(m_\phi^2-\omega_0^2)^2}\frac{\dd U}{\dd f}-
\left[\Omega-\alpha A(R^\ast)\right]^2f \,.\label{eq:fEOM}
\eeq
 Equation~\eqref{eq:fEOM} is exactly the form of the equation for the \emph{global} Q-ball 
Eq.~\eqref{e.fGlobal} with the global value of $\Omega_G$ given by
\beq
\Omega_G=\Omega-\alpha A(R^\ast)\,.\label{e.gaugeOm12}
\eeq

Since the derivative of $\alpha A$ is small, it does not contribute significantly to 
the friction over the transition region. This means that the frictional effects over the 
transition are also nearly identical to the global case.  Since the relation between $\Omega_G$ and $R^\ast$ is determined by the friction, 
if the $R^\ast$ dependence of the global Q-ball parameter $\Omega_G(R^\ast)$ is known, we 
can determine the $R^\ast$ dependence of the gauged Q-ball $\Omega(R^\ast)$ via
\begin{align}
\boxed{
\Omega(R^\ast)= \Omega_G(R^\ast)\,\alpha R^\ast \coth(\alpha R^\ast) \, ,}\label{e.gaugeOm}
\end{align}
where we have used Eq.~\eqref{e.gaugeOm12} and the thin-wall formula of Eq.~\eqref{e.thinA} for $A(R^\ast)$.

Equation~\eqref{e.gaugeOm} is the key result of our article. It provides a mapping from global 
Q-balls\textemdash for which the relation $\Omega_G(R^\ast)$  is 
much easier to obtain both analytically and numerically\textemdash and gauged Q-balls with any $\alpha$. 
 Furthermore, the scalar transition profiles for the 
{gauged} Q-balls are expected to be identical to the transition profiles for the corresponding 
{global} Q-balls (Eq.~\eqref{eq:transition_profile}). 
As we now show, this rather simple argument leads to  accurate analytic descriptions of gauged 
Q-balls.

%%%%%%%%%%%%%%%%%%%%%%%
%%%%%%%%%%%%%%%%%%%%%%%
\section{Results}
\label{sec:results}

{ We can now use these results to construct an analytical estimate for the Q-ball profile}.
The mapping in Eq.~\eqref{e.gaugeOm} provides the radius of the gauged Q-ball given the known 
relationship $\Omega_G (R^\ast)$ from the global Q-ball 
(Eq.~\ref{e.globalRapproximation}). 
The scalar profile $f(\rho)$ is taken to be the transition profile of global Q-balls 
(Eq.~\eqref{eq:transition_profile}); this is well motivated around $\rho\sim R^\ast$ 
for large $R^\ast$ but happens to be a very good approximation for all other cases as 
well. Finally, the gauge profile $A(\rho)$ is taken from Eq.~\eqref{e.thinA}.
We can also use this analytical profile to find approximations 
for $Q$ (via Eq.~\eqref{e.charge}) and $E$ (via Eq.~\eqref{e.energy}); 
since the resulting expressions are lengthy we do not show them here.

\begin{figure}[t]
\includegraphics[width=0.49\textwidth]{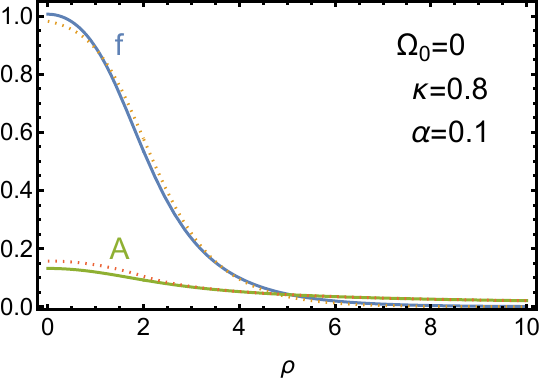}
\includegraphics[width=0.49\textwidth]{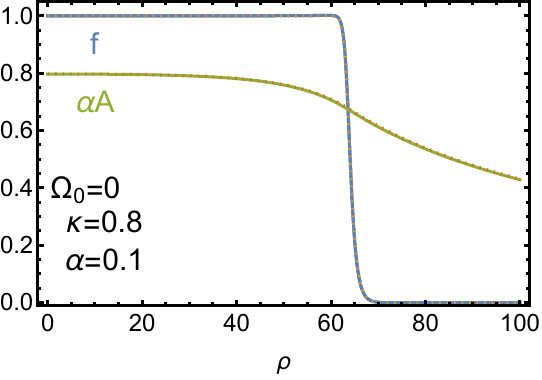}
\caption{Profiles for the scalar field and gauge field for a thick-wall (left) Q-ball and a thin-wall (right) 
Q-ball. The exact numerical results are denoted by the solid lines, while the thin-wall analytic approximation 
is given by the dashed lines. }
\label{fig:profilePlot}
\end{figure}

These profiles serve as excellent seed functions for the numerical solution of the differential equations 
described in Sec.~\ref{sec:numerics}.
Figure~\ref{fig:profilePlot} shows a
comparison between the numerical calculations and our analytical estimates
for one choice of parameters.
 Note that the two solutions in Fig.~\ref{fig:profilePlot} have the same potential 
parameters and scalar frequency $\omega$, but differ in their Q-ball observables 
such as radius, charge, and energy. These two solutions correspond to the two solutions for $R^\ast$ obtained from the mapping in Eq.~\eqref{e.gaugeOm}.
As  
the plot illustrates, the analytical profiles for $f$ and $A$ match the numerical results remarkably well, especially for the 
large  Q-balls (right panel).

\begin{figure}[t]
\includegraphics[width=0.485\textwidth]{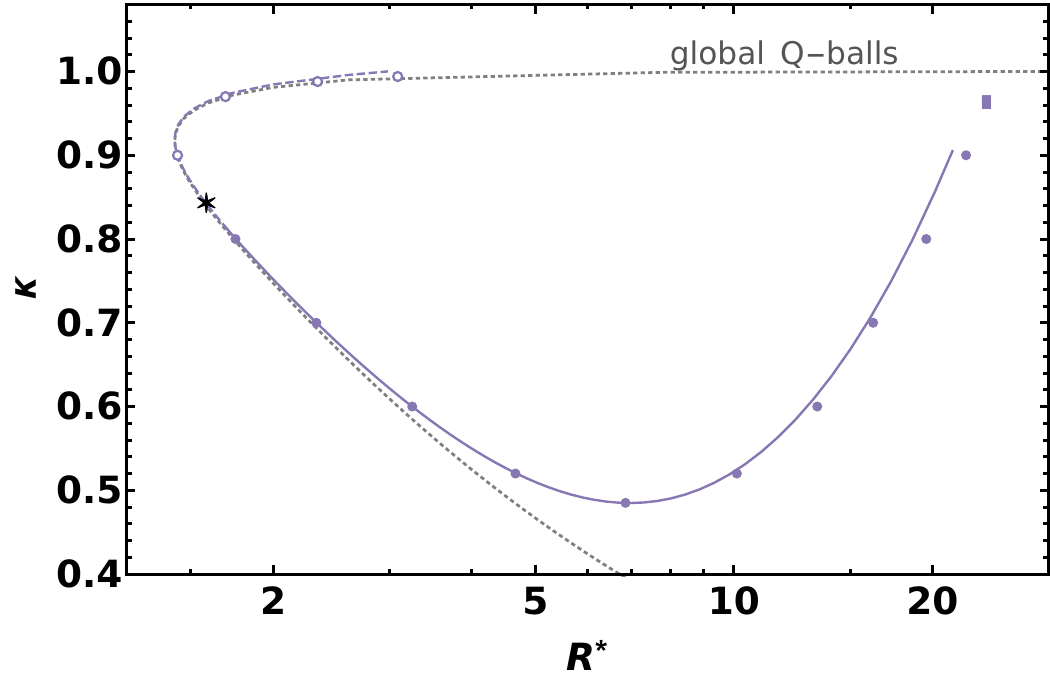}
\includegraphics[width=0.495\textwidth]{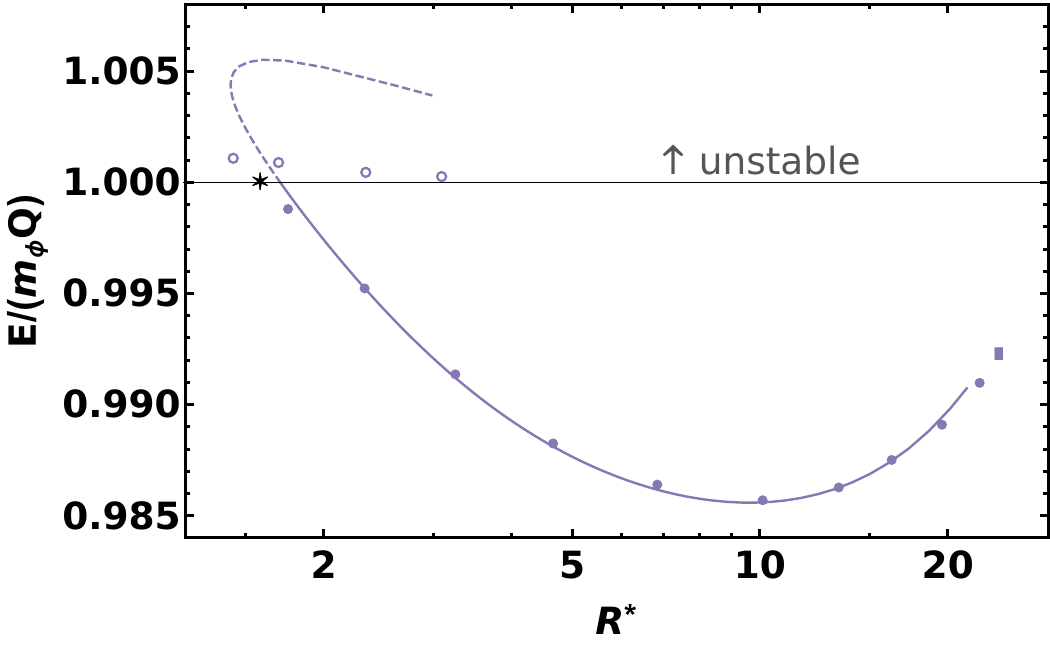}
\includegraphics[width=0.49\textwidth]{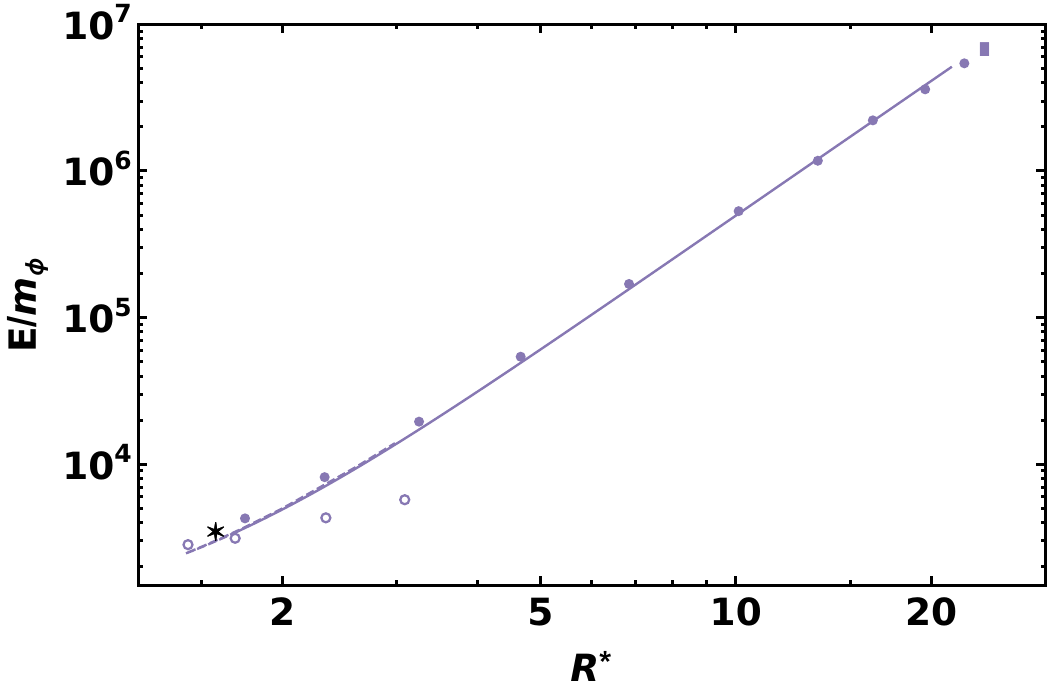}
\includegraphics[width=0.49\textwidth]{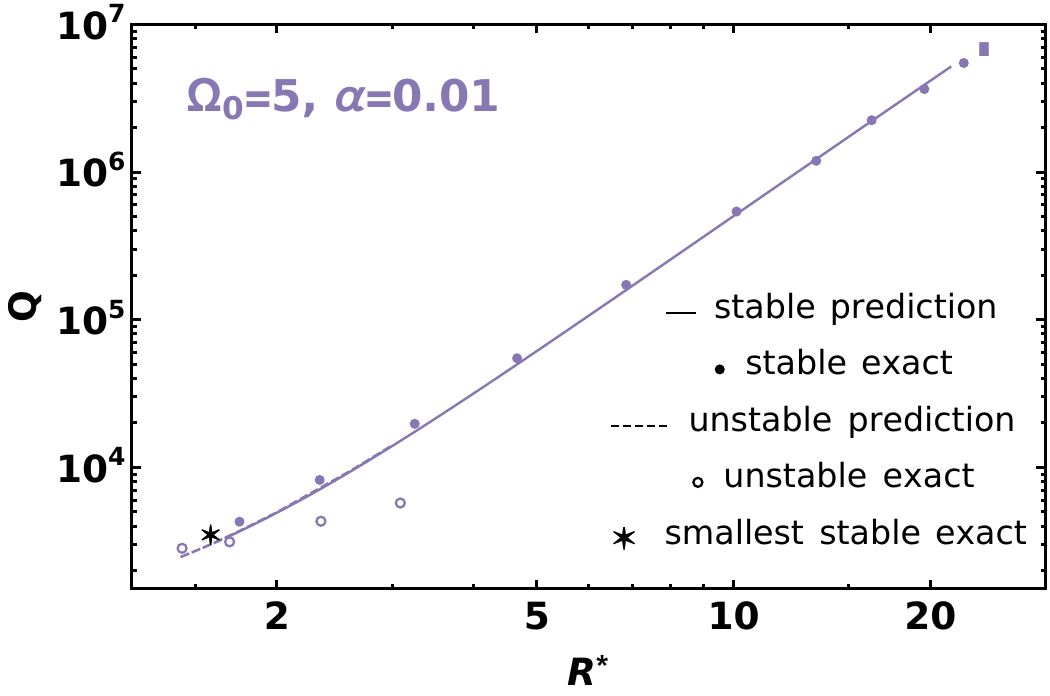}
\caption{
A comparison of predictions from Eq.~\eqref{e.gaugeOm} and numerical solutions 
for a sample benchmark $\Omega_0=5$, $\alpha=0.01$, $\phi_0 = m_\phi$.
Predicted stable and unstable solutions are shown as solid and dashed lines, respectively,   
and stable and unstable numerical solutions are shown as filled and open points, respectively.  
The gray dotted line  shows the global Q-ball case. The rectangle shows the largest numerical solution. }
\label{fig:benchx}
\end{figure}

We now discuss the Q-ball observables for 
the benchmark point $\Omega_0 = 5$ and $\alpha = 1/100$; we set 
$\phi_0 = m_\phi$ throughout and measure all dimensional quantities in units of $m_\phi$.
The results for this benchmark are shown in Fig.~\ref{fig:benchx}.
In the top left panel, 
the numerical results for
 $\kappa$ vs.~$R^\ast$ (circles) are compared with
the prediction obtained from Eq.~\eqref{e.gaugeOm} (line). The other panels 
show the analogous results for $E/m_\phi, Q$, and  $(E/m_\phi Q)$.
Overall, there is excellent agreement between the numerical and analytical results.

There are a number of features that restrict the allowable Q-ball solutions. First, we
must have  $\omega\leq m_\phi$ (or $\kappa \leq 1$) in order for the Q-ball solution to relax to zero 
for large $\rho$. This typically\footnote{The functional form of $\Omega_G (R^\ast)$ depends on the scalar potential. 
Equation~\eqref{e.gaugeOm} implies that $\Omega_G (R^\ast)$ must fall off faster than $1/R^\ast$ at 
large $R^\ast$ in order to construct gauged Q-balls without a maximal radius. We are not aware of such 
potentials and global Q-balls in the literature.}  implies a maximum Q-ball radius.
Secondly, we must have $E \leq m_\phi Q$ so that the Q-ball is stable against decay to scalars.
This constraint is most easily seen in the top right panel of Fig.~\ref{fig:benchx},
and 
implies the existence of a \emph{minimal} Q-ball radius. We have shown this second 
instability by representing our prediction by a dashed line
in the unstable region. The numerical solutions show the same instability; we have represented the
last stable solution (the stable solution with smallest $R^\ast$) as a star.

Finally, we must impose the constraint of Eq.~\eqref{e.ArvindsInequality} that demands that
the scalar potential have a second maximum away from $f=0$. 
This puts an upper bound on the radius which, for this benchmark,
is more restrictive than the 
maximal radius determined by the relation $\omega \leq m_\phi$.
Using Eq.~\eqref{e.ArvindsInequality} with $A = A(0)$ from our thin-wall 
expression Eq.~\eqref{e.thinA},
we can calculate this maximal radius $R^\ast_\text{max}$ and impose this constraint on our analytical prediction 
shown in the figure, ending the solid line before $\kappa=1$ .
Since the thin-wall $A(0)$ overestimates the true value, our maximal radius is slightly smaller 
than the true maximal radius (indicated by a rectangle 
in the plot), but the  agreement is still good.

One interesting feature in the $\kappa$ vs.~$R^\ast$ plot 
is the existence of a minimum allowed value of $\kappa$.
An analytic 
expression for this minimum value
can be obtained; since this minimum value must be less than or equal to one
for Q-balls to exist, we find the constraint
\begin{align}
\alpha \lesssim \frac{1}{\sqrt{1/(0.58)^2 + 9 \Omega_0^2/2}} \,.
\end{align}
In particular, this predicts that there are no gauged Q-balls with $\alpha > 0.58$.
Numerically, we find that the actual upper limit for $\alpha$ is $0.52$, in quantitative 
agreement with the above mapping derivation. 
Note that it was pointed out in 
Ref.~\cite{Lee:1988ag}
that for any scalar potential (and its implied attractive force) 
there must be an upper bound on the allowed gauge coupling (and its implied repulsive force) 
in order to form a stable Q-ball.

\begin{figure}[t]
\includegraphics[width=0.49\textwidth]{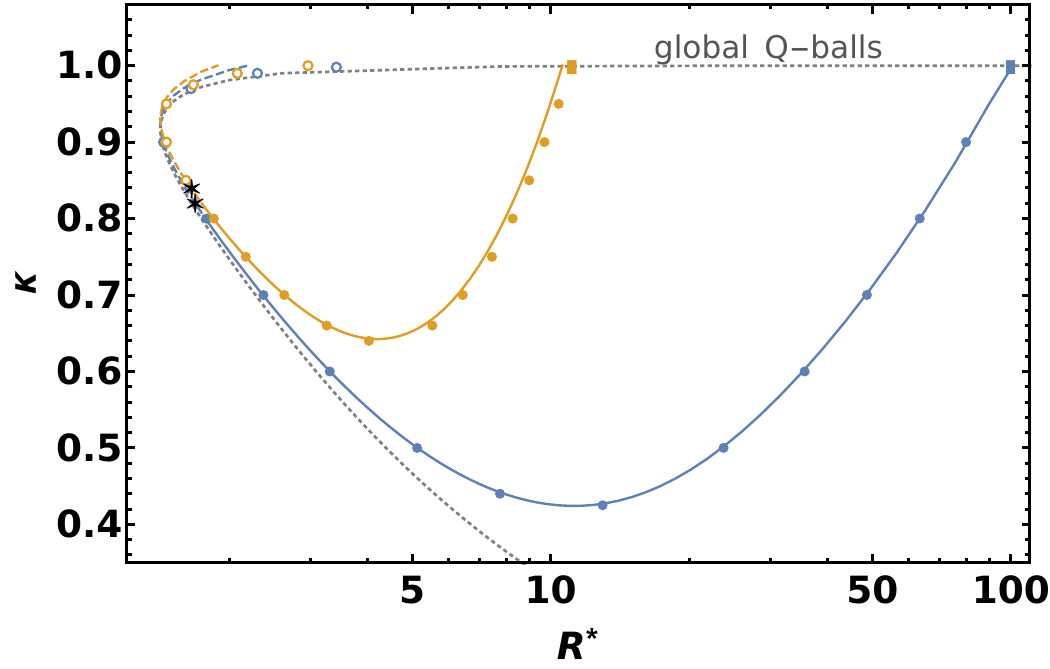}
\includegraphics[width=0.495\textwidth]{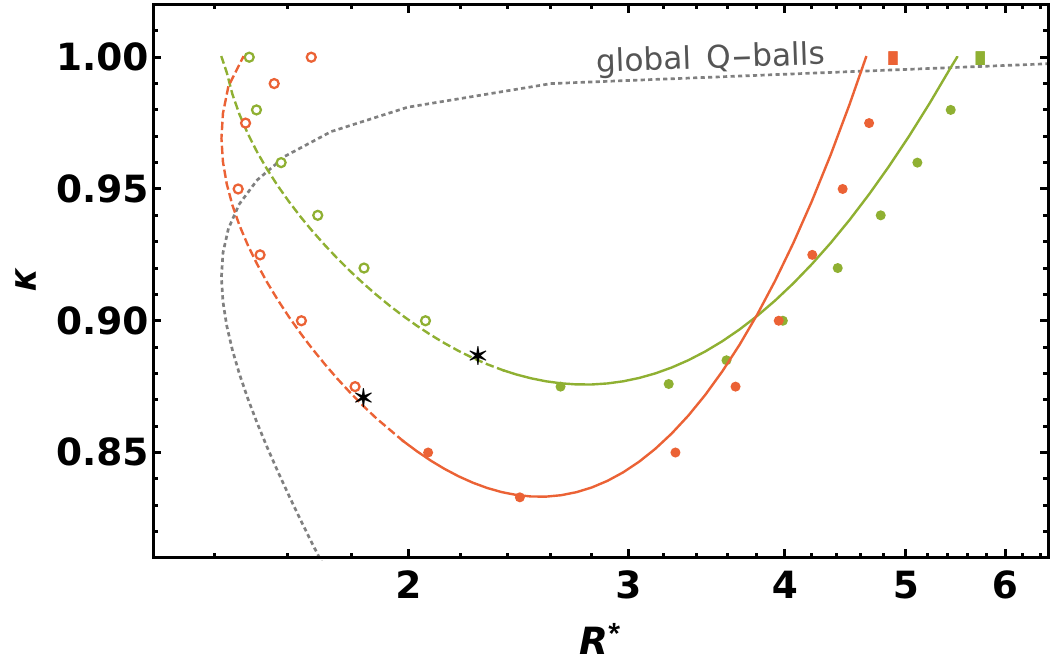}
\includegraphics[width=0.48\textwidth]{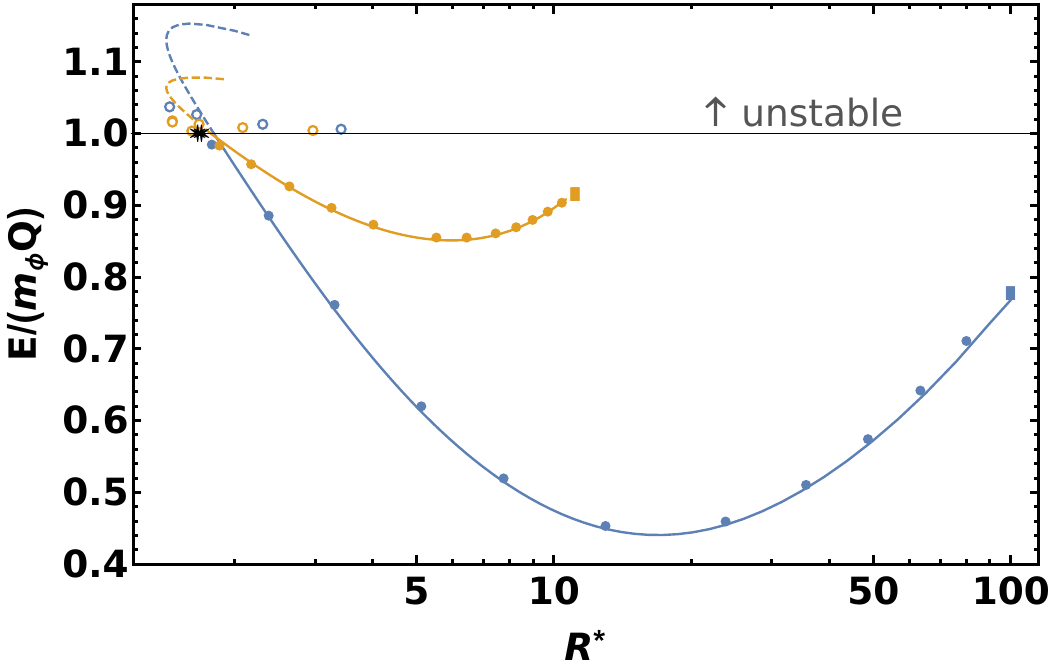}
\includegraphics[width=0.49\textwidth]{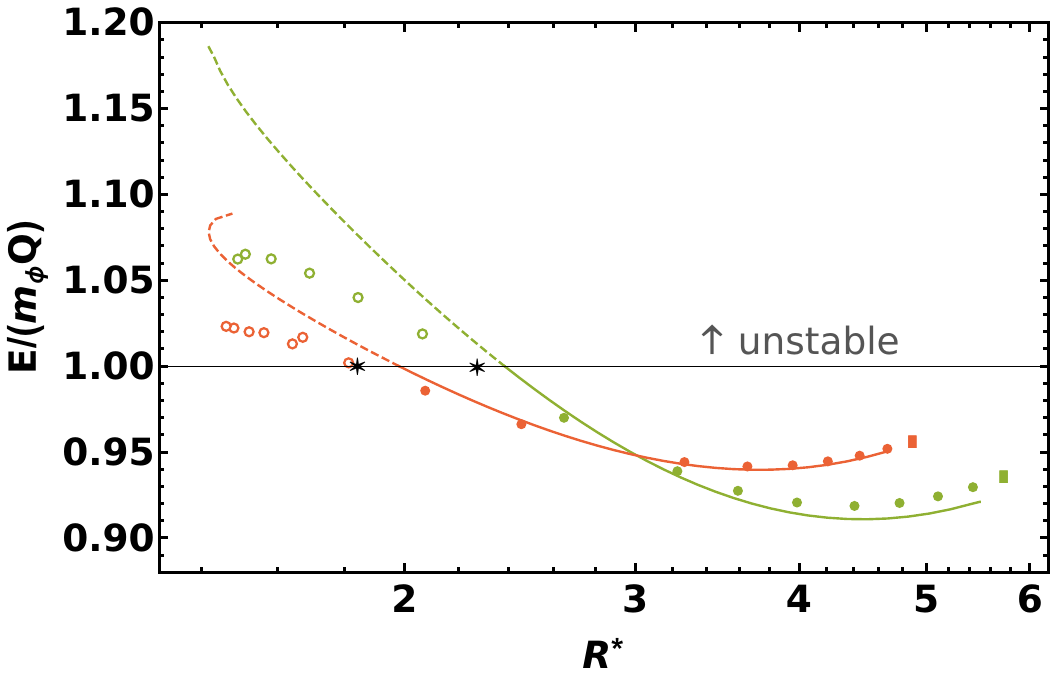}
\includegraphics[width=0.49\textwidth]{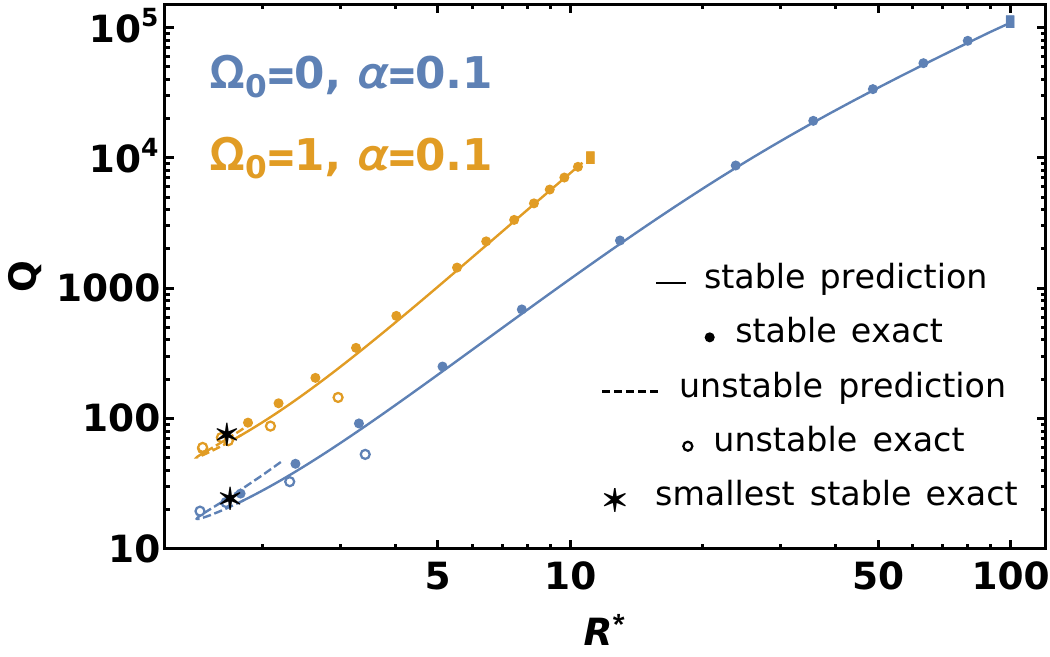}
\includegraphics[width=0.49\textwidth]{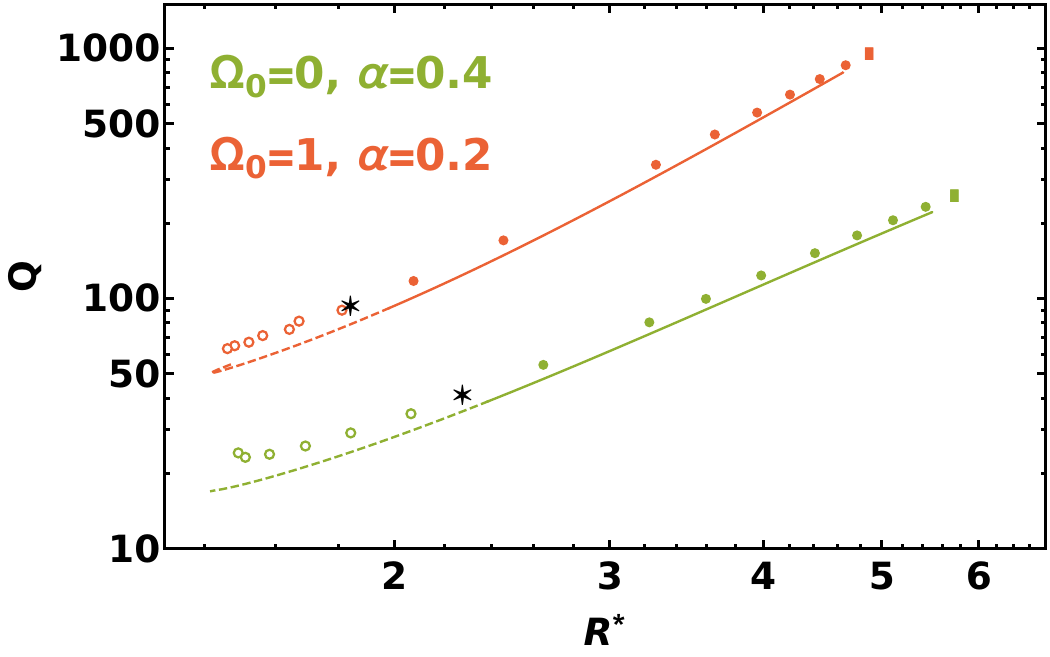}
\caption{A comparison of predictions from Eq.~\eqref{e.gaugeOm} and numerical solutions for 
benchmarks $\Omega_0=0$, $\alpha=0.1$ (left, blue), $\Omega_0=1$, $\alpha=0.1$ (left, orange).   
$\Omega_0=0$, $\alpha=0.4$ (right, green) and $\Omega_0=1$, $\alpha=0.2$ (right, red).  
Conventions are as in Fig.~\ref{fig:benchx}.  }
\label{fig:bench1to4}
\end{figure}
%\pagebreak

The lower panels of Fig.~\ref{fig:benchx} show the behavior of $Q$ and $E$ as a function of
~$R^\ast$. They  inherit both a 
minimal and a maximal value from the corresponding radius. Our analytical 
predictions match 
the numerical results on the (phenomenologically interesting) stable Q-ball branch.

We compare the analytical and numerical data for several other benchmarks
 in Fig.~\ref{fig:bench1to4}. Our predictions 
show only small deviations with 
respect to the numerical results for all benchmarks. This illustrates that the mapping in Eq.~\eqref{e.gaugeOm} holds qualitatively and quantitatively over the whole parameter space.

For these benchmarks,   $R^\ast_\text{max}$ is set by the condition 
$\kappa =1$ rather than by
 Eq.~\eqref{e.ArvindsInequality}.
Using Eq.~\eqref{e.gaugeOm} and the large-$R^\ast$ relation $\kappa_G = 1/\sqrt{R^\ast}$~\cite{Heeck:2020bau} we find
\begin{align}
 \alpha^2 R^\ast_\text{max} (1 + R^\ast_\text{max} \Omega_0^2) \coth^2 (R^\ast_\text{max} \alpha) - \Omega_0^2 = 1\,.
\end{align}
This equation cannot be solved analytically, but has the limiting cases:
\begin{align}
R^\ast_\text{max} \simeq 
\begin{cases}
\frac{1}{\alpha^2} \,, & \text{ for } \Omega_0 \lesssim \alpha\,,\\
\frac{1}{\alpha \Omega_0} \,, & \text{ for } \Omega_0 \gg \alpha\,.
\end{cases}
\end{align}
Since both charge and energy grow with $R^\ast$ for large radii, this $R^\ast_\text{max}$ also implies a 
maximal Q-ball charge and energy for a given set of potential parameters. This qualitative claim was made in Ref.~\cite{Lee:1988ag}, but here we provide easy-to-use quantitative predictions.

We also note that in the limiting situation of large $R^\ast$, 
the expressions for charge 
and energy 
simplify to
\begin{align}
Q &= \frac{4 \pi}{\alpha ^2} \Phi_0^2  \left(\alpha  R^\ast \coth (\alpha   R^\ast)-1\right)  
\sqrt{ {R^\ast}^2 \Omega_0^2+R^\ast} \,,\label{eq:thinQ} \\
E &= \frac{\pi  m_\phi  R^\ast \Phi_0^2 \text{csch}^2(\alpha  R^\ast) }{6 \alpha  \sqrt{\Omega_0^2+1}}
\left[\alpha   R^\ast \left(4  R^\ast \Omega_0^2+3\right)-6 \left( R^\ast \Omega_0^2+1\right) 
\sinh (2 \alpha  R^\ast) \right.\nonumber\\
&\qquad\qquad\qquad\qquad\qquad\qquad\left.+\alpha  R^\ast \left(8 R^\ast \Omega_0^2+9\right) 
\cosh (2 \alpha  R^\ast)\right] ,\label{eq:thinE}
\end{align}
as derived in App.~\ref{sec:alternative_derivation}.
These are more approximate than the full integrals used in our figures, but are significantly more 
manageable and still make excellent predictions at large $R^\ast$.

Using our analytical approximations together with numerical results, we can show 
that stable gauged Q-balls have 
$R^\ast \gtrsim 1.5$, which is similar to the lower limit 
found for global Q-balls~\cite{Heeck:2020bau}. This matches the physical 
expectation that the introduction of a repulsive force to a Q-ball should not decrease the 
Q-ball radius.

We note that for $\Omega_0=0$, the scalar profile is found to be essentially constant in the interior of
thin-wall Q-balls (Fig.~\ref{fig:profilePlot}, right), and our approximations become more accurate, especially for small $\alpha$, where the solutions approach the global Q-ball case. For larger $\Omega_0$, the solutions deviate from the global case (Fig.~\ref{fig:potentialPlot}, left), but our results remain accurate. It would be
interesting to explore the dependence on $\Omega_0$ further; we leave this to future work.

\section{Conclusion}
\label{sec:conclusion}

Global Q-balls are curious objects that arise in certain $U(1)$-symmetric scalar field 
theories and can be studied analytically and numerically with relative ease. Promoting the 
$U(1)$ symmetry to a gauge symmetry complicates the discussion significantly and has eluded 
analytical descriptions outside of some limiting cases. 

In this article we have exhibited a method
to obtain essentially all properties of gauged 
Q-balls via a mapping from global Q-balls. Since the latter can be easily 
obtained numerically and often even analytically, this mapping allows for an excellent 
prediction of the gauged Q-ball properties without the need to solve the coupled, nonlinear 
differential equations. 
Our analytical expressions also make possible the solution of the differential 
equations by finite-element methods rather than the shooting method.

Finally, we stress that our analytical approximations are best in the thin-wall or 
large-radius limit.
Smaller Q-balls show 
larger deviations, but these are also the Q-balls that are easiest to study numerically, 
providing good complementarity. Importantly, our analytical approximations also serve as 
good seed functions for numerical finite-element methods, significantly simplifying the 
numerical study of thick-wall gauged Q-balls.

\section*{Acknowledgements}
This work was supported in part by NSF Grant No.~PHY-1915005. C.~B.~V.~also acknowledges support from Simons 
Investigator Award \#376204. R.~R. acknowledges support from the National Science Foundation Graduate 
Research Fellowship Program under Grant \#1839285.

\appendix
\section{Energy of Gauged Q-Balls\label{a.Energy}}
In this appendix we derive the form of the energy given in Eq~\eqref{e.energy}. We begin with the Lagrangian and rescale the radial coordinate $\rho\to\chi \rho$. This yields
\begin{equation}
L=4\pi\Phi_0^2\sqrt{m_\phi^2-\omega_0^2}\int \dd\rho\,\rho^2\chi\left[ -\frac12f^{\prime 2}+\frac{1}{2}A^{\prime2}+\chi^2V(f,A)\right],
\end{equation}
where $V(f,A)$ is defined in Eq.~\eqref{eq:potentialV}. We now consider the variation of the Lagrangian with respect to $\chi$ and then set $\chi=1$. The variation has two parts, first the explicit dependence on $\chi$ and second the variation that appears because functions now depend on $\chi$, $f(\rho)\to f(\rho\chi)$. This second collection of terms, with $\chi$ then set to one, is simply the usual variation of the Lagrangian, and so vanishes by definition. Requiring the other term in the variation to also vanish yields the constraint
\begin{equation}
0=\int \dd\rho\,\rho^2\left[ -\frac12f^{\prime 2}+\frac{1}{2}A^{\prime2}+3V(f,A)\right].\label{e.Lconst}
\end{equation}

We can use this constraint to remove the explicit dependence on $U(f)$ from the energy in Eq.~\eqref{e.Eint}:
\begin{align}
E=&4\pi\Phi_0^2\sqrt{m_\phi^2-\omega_0^2}\int \dd\rho\,\rho^2\left[\frac13f^{\prime 2}+\frac{2}{3}A^{\prime2} +f^2(\alpha A-\Omega)^2 \right]\nonumber\\
=&4\pi\Phi_0^2\sqrt{m_\phi^2-\omega_0^2}\int \dd\rho\,\rho^2\left[\frac13f^{\prime 2}+\frac{2}{3}A^{\prime2}+\frac{1}{\alpha\rho^2} (\alpha A-\Omega)\left(\rho^2A' \right)'\right],
\end{align}
where in the last line we have used the $A$ equation of motion in \eqref{e.Aeq}. The third term is then integrated by parts to produce
\begin{align}
\frac{E}{\sqrt{m_\phi^2-\omega_0^2}}=&\Omega Q+\frac{4\pi\Phi_0^2}{3}\int \dd\rho\,\rho^2\left(f^{\prime 2}-A^{\prime2} \right)~.\label{e.EnoU}
\end{align}
This result is useful in that it only depends on the change in $f$ and $A$. Alternatively, the form
\begin{align}
\frac{E}{\sqrt{m_\phi^2-\omega_0^2}}=\Omega Q+8\pi\Phi_0^2\int \dd\rho\,\rho^2V(f,A)~,
\end{align}
can also be used to determine the energy without any use of the derivatives of $f$ and $A$.

\section{An Alternative Mapping Derivation}
\label{sec:alternative_derivation}

As an alternative to the derivation of the mapping equation~\eqref{e.gaugeOm} in 
Sec.~\ref{sec:Mapping} we provide here a derivation in the thin-wall limit, i.e.~for 
large $R^\ast$. For this we consider the simplest thin-wall ansatz for the 
profiles~\cite{Lee:1988ag}, where $f$ is a step function, 
$f(\rho) \simeq 1 - \Theta(\rho-R^\ast)$, and $A$ is given by Eq.~\eqref{e.thinA}. We can 
easily integrate these functions to obtain the charge $Q$~\cite{Lee:1988ag},
 \begin{align}
 Q = \frac{4 \pi  \Omega  \Phi_0^2 }{\alpha ^3 } \left(\alpha  R^\ast-\tanh (\alpha  R^\ast)\right) ,
 \end{align}
and the energy [Eq.~\eqref{e.energy}], 
 \begin{align}
E = \omega Q 
+\frac{\pi \phi_0^2 }{3\sqrt{m_\phi^2 - \omega_0^2}} {R^\ast}^2
-\frac{4\pi \phi_0^2 }{3 \sqrt{m_\phi^2-\omega_0^2}} \, \frac{\Omega ^2 \left(\alpha  R^\ast \left(\text{sech}^2(\alpha  R^\ast)+2\right)-3 \tanh (\alpha  R^\ast)\right)}{2 \alpha ^3} \,,
\end{align}
using the results from Ref.~\cite{Heeck:2020bau} to properly integrate over the discontinuous $f'^2$.
Notice that the last term in $E$ goes to zero for $\alpha\to 0$, leading back to the global case. 
Now we can use the equation~\eqref{e.dedw} in the form $\dd E/\dd R^\ast = \omega(R^\ast) \dd Q/\dd R^\ast$ to obtain -- and solve -- a differential equation for $\omega(R^\ast)$, yielding
\begin{align}
\omega(R^\ast) = \coth (\alpha  R^\ast)  \sqrt{c {R^\ast}^2+\alpha ^2 (m_\phi^2- \omega_0^2)R^\ast} \,.
\end{align}
Here, $c$ is an integration constant that is difficult to obtain, but we can get $c = \alpha^2 \omega_0^2 + \mathcal{O}(\alpha^3)$ from matching to the global case $\kappa^2 \simeq 1/R^\ast$ (valid roughly for $R^\ast > 2$). This gives us
\begin{align}
\omega(R^\ast) = \alpha R^\ast \coth (\alpha  R^\ast) \sqrt{\omega_0^2+ \frac{m_\phi^2- \omega_0^2}{R^\ast}} \,,
\end{align}
which is identical to the more general mapping formula in Eq.~\eqref{e.gaugeOm} in the large $R^\ast$ limit due to  $\omega_G\simeq \sqrt{\omega_0^2+ \frac{m_\phi^2- \omega_0^2}{R^\ast}}$~\cite{Heeck:2020bau}.

\bibliographystyle{utcaps_mod}
\bibliography{BIB}

\end{document}